# Potassium L-ascorbate monohydrate: a new metal-organic nonlinear optical crystal

Dhanpal Bairwa[1*], K Raghavendra Rao[2], Diptikanta Swain[3], T.N. Guru Row[3], H.L. Bhat[1], Suja Elizabeth[1]

[1] Department of physics, Indian Institute of Science, Bangalore, India, 560012
[2] PES University, Bangalore, India, 560085
[3] Solid State and Structural Chemistry Unit, Indian Institute of Science, Bangalore, India, 560012

Corresponding author: dhanpal@iisc.ac.in Tel: +91 (80) 22932721

## Abstract

Large size single crystals of potassium L-ascorbate monohydrate (KLAM), ($KC_6H_7O_6.H_2O$) are grown using solution growth technique by lowering the temperature at the rate of 0.24 °C/h, where water was used as solvent. The structure of KLAM was solved by single crystal XRD. KLAM crystallizes in non-centrosymmetric, monoclinic, $P2_1$ space group with lattice parameters a = 7.030(5) Å, b = 8.811(5) Å, c = 7.638(5) Å and β = 114.891(5)°. The crystal grows with bulky morphology in all three directions having (100), (-100), (-110), (0-1-1), (0-11), (001) and (00-1) prominent faces. TGA and DSC measurements show that KLAM is stable up to 80 °C. The crystal shows good optical transparency with a lower cut off as low as 297 nm. Second harmonic conversion efficiency measured on powder sample is 3.5 times that of potassium dihydrogen phosphate (KDP). Phase matching (PM) is observed on a plate of the KLAM. Noncollinear phase matching rings are also observed near the PM directions which help to identify the locus of PM directions. Presence of noncollinear SHG rings up to third order suggests large birefringence and nonlinear optical coefficients. Laser damage threshold value of the crystal is found to be 3.07 $GW/cm^2$, at 1064 nm in 100 direction.

## Introduction

Nonlinear optical crystals play key role in solid state lasers [1-4], which emit light at varying frequencies through nonlinear optical processes [5-14]. Due to their versatile applications in scientific and technological applications, they have received research extensive interest. Nonlinear optical crystals have advantages of low cost, chemical stability, large second harmonic coefficients, moderate birefringence ($\Delta n \sim 0.05-0.10$), wide transparency window with high transmission and high laser damage threshold all of which are of scientific and technological interest [15-17]. Saccharides (sugars) are organic materials composed of carbon, hydrogen and oxygen. The molecules of saccharides are chiral and crystallize in non-centrosymmetric crystal structure which is a prerequisite for generation of second harmonic [5]. L-ascorbic acid is a monosaccharide which is water soluble and forms colorless bulky crystals due to multi-directional hydrogen bonding [18]. Saccharides have higher melting point and are harder than van der wall bonded materials [19], but crystals which include water molecules in their structure lose crystallinity when water goes out. Saccharide crystals possess second harmonic generation (SHG) capability and are potential material for phase matching [20]. Previous reports on lithium l-ascorbate dihydrate, lithium D-isoascorbate monohydrate, sodium D-isoascorbate monohydrate, D-isoascorbic acid have shown that saccharides indeed possess moderate to large birefringence and low dispersion and therefore are easily phase-matchable [21-24]. Ascorbic acid and its salts are not very stable in solution as they oxidize and degrade however they are quite stable in crystalline state [25]. A previous report on lithium salt of ascorbic acid shows that the introduction of lithium ion improves its chemical and physical properties. Besides, crystal structure is modified which leads to enhancement in nonlinear optical properties like conversion efficiency, laser damage threshold, and stability [23]. For investigating the effect of other alkali metals like potassium, we synthesized the potassium salt of l- ascorbic acid, which is monohydrate, crystallizes in monoclinic $P2_1$ noncentrosymmetric space group which satisfies the structural requirement for second harmonic generation. In this paper, we discuss the synthesis, crystal growth, structural and thermal (TGA and DSC) properties of potassium l- ascorbate monohydrate KLAM. Also, powder SHG by Kurtz Perry method, noncollinear and collinear phase matching and laser damage are investigated.

## Experimental section

KLAM was synthesized by mixing L-ascorbic acid (99.99%, Sigma Aldrich) and potassium carbonate (99.99%, Sigma Aldrich) in stoichiometric ratio (2:1) in distilled water at room temperature. L-ascorbic acid reacts with potassium carbonate quickly with evolution of $CO_2$ . The chemical reaction is given below.

$$2C_6H_8O_6 + K_2CO_3 + H_2O \rightarrow 2\ KC_6H_7O_6.H_2O + CO_2\uparrow$$

The solution is stirred using a magnetic stirrer until the evolution of $CO_2$ complete. The solution was avoided from light and temperature exposure as it is sensitive to them and unstable in these conditions [25]. The color of the solution changes from clear to yellowish-red over a period of time due to photochemical decomposition [26].

Before growing the crystal, the compound was recrystallized twice to increase purity. Crystals up to the size of 55 × 30× 15 $mm^3$ were grown without seeding, in about 4 weeks. Solution growth technique using modified Holden’s rotary crystallizer yields full morphology and hence employed for this work. It was observed that solubility of the material increases with temperature, so crystals were grown by lowering the temperature in the range of 33-28 °C. The upper limit of the temperature of the saturated solution was kept at 33°C keeping in mind the solution’s sensitivity to temperature. Seed crystals were first obtained by slow evaporation of water. A good crystal thus obtained was introduced as seed in the saturated solution which was cooled at the rate of 0.01 °C/h using Eurotherm 2416 temperature controller. After about 15-20 days, crystals of size 35× 35× 20 $mm^3$ in size were obtained. The photographs of as grown crystals are shown in Figure.1.

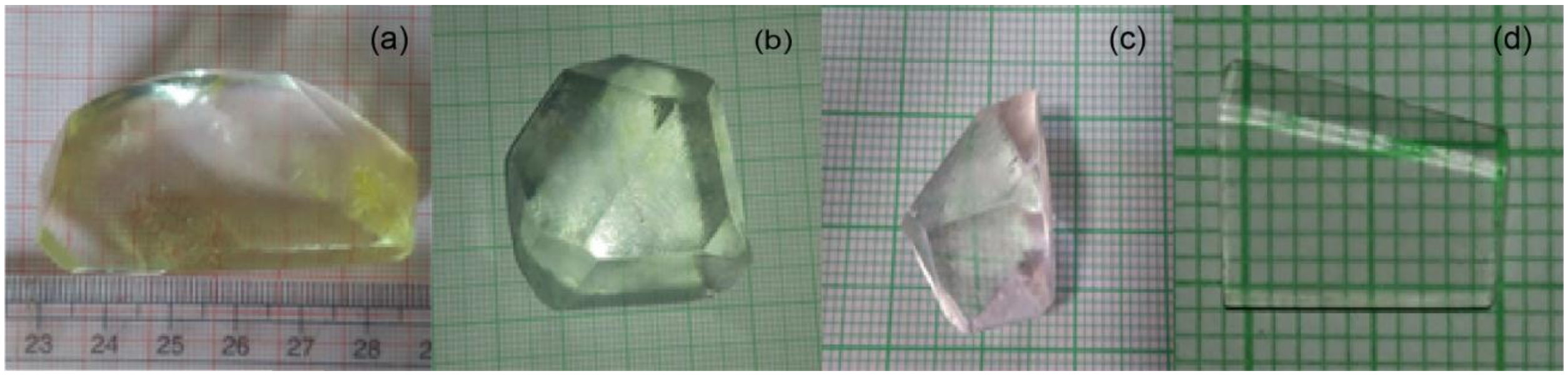

Figure.1. (a) A KLAM crystal grown by slow evaporation without seeding. (b) and (c): crystals grown with solution growth set up by seeding. (d) Polished b plate

Single crystal X-ray diffraction (SCXRD) was performed using Oxford Xcalibur diffractometer with Eos detector and Mova microscope (Mo-$K_\alpha$ radiation, $\lambda$ = 0.71073 Å). The structure was solved by direct method in Shelx2014. To check the bulk purity and presence of single phase in the entire crystal, powder X-ray diffraction was carried on Rigaku SmartLab X-ray diffractometer which employs $CuK_\alpha$ radiation. The interfacial angles were measured using a contact goniometer with 0.1° least count. Thermogravimetric analysis (TGA) was made with the help of ME-51140728 model of ARTESYN technologies while Differential scanning calorimetry (DSC) experiments were carried out with the help of DSC 2920 model of TA Instruments. The UV-Vis transmission spectrum was recorded using UV-3600 spectrometer. For all optical measurements cut and polished (using methanol and $Al_2O_3$ powder) crystal plates were used. For polishing, the crystal plates were rubbed on the BUEHLER's polishing micro cloth with methanol and alumina ($Al_2O_3$) powder of different particle size (3-0.5μm). Crystals were cut using a homemade wet-saw machine, which cuts the crystal by dissolving the crystal by a fine wet string. An Nd:YAG laser (Brilliant B, Quantal, 1064 nm, 10 Hz repetition rate, 6 nm pulse width) was used for collinear and noncollinear phase matching. For laser damage studies, Nd:YAG and He-Ne lasers were used simultaneously as pump and probe. The energy of Nd:YAG and He-Ne lasers were detected using pyroelectric sensors PE25BBDIFSHV2 (with and without diffuser) and 3APSHV1, respectively together with a NOVA II Ophir power meter. For visual verification of the damage, a Leitz Orthoplan microscope was used. For the relationship between crystallographic a, b, c directions and the direction of optical importance, Laue XRD was carried out using Rigaku diffractometer with Mo-$K_\alpha$ radiation (polychromatic). Optical measurements were carried out by keeping the crystal on goniometer. The goniometer with the crystal was then transferred to X-ray machine gently so as not to disturb the crystal from its position. Data were fitted using Orient express software. Rocking curve measurement was carried out using Rigaku SmartLab X-ray diffractometer.

# Results and discussion

## Single crystal XRD

Crystallographic information obtained from SCXRD and details of data refinement are given in Table I. KLAM crystallizes in monoclinic $P2_1$ space group at room temperature. Molecular diagram of KLAM is given in Figure.2. Along the **b** direction potassium atoms are connected via O3 and O7W atoms. In **a** and **c** directions they are connected through organic links as shown in the packing diagram of KLAM in Figure.3.

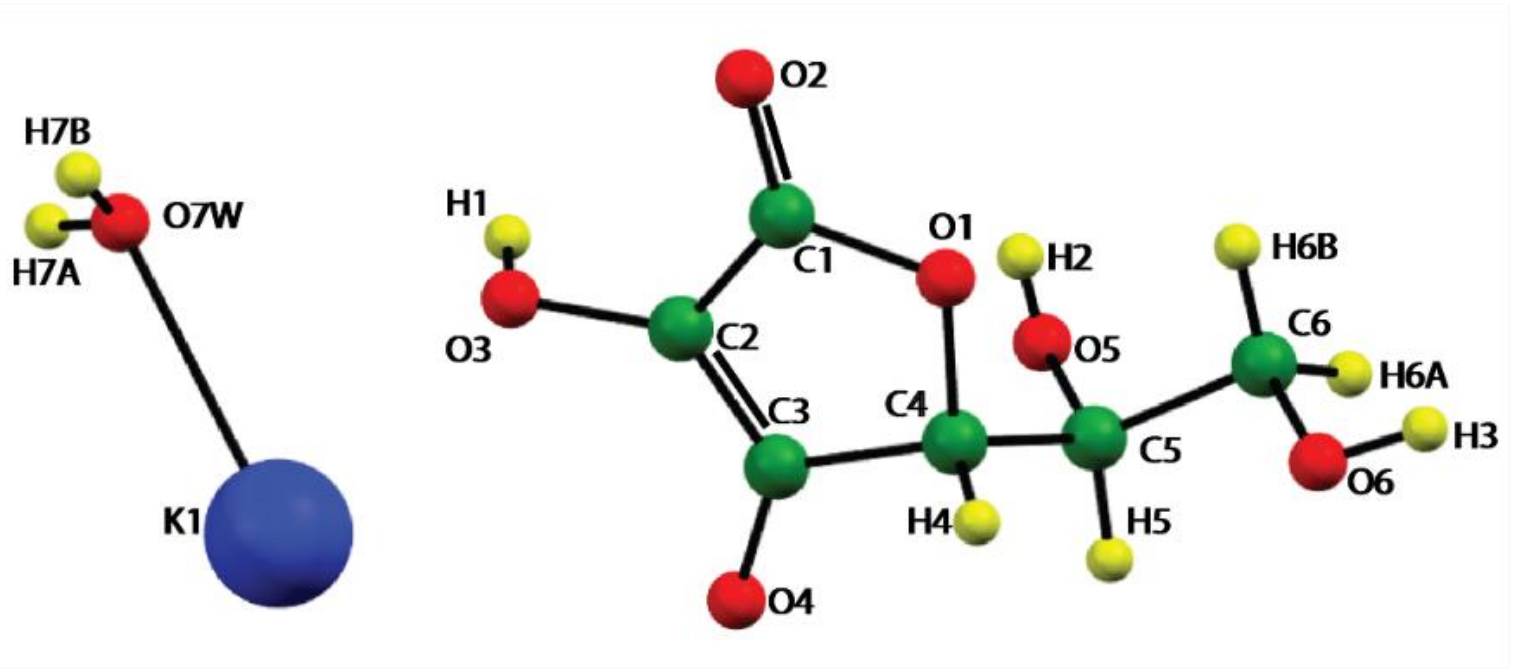

Figure.2. Molecular diagram of potassium l-ascorbate monohydrate.

Table I. Crystallographic data of KLAM obtained from SCXRD.

| | |
|---|---|
| Formula | $KC_6H_7O_6.H_2O$ |
| Formula weight | 232.23 |
| Temperature (K) | 295K |
| Colour | Colourless |
| Crystal size (mm) | 0.2×0.15×0.12 |
| Radiation | Mo $K_\alpha$ |
| Wavelength (Å) | 0.7107 |
| Crystal system | Monoclinic |
| Space group | $P2_1$ |
| *a* (Å) | 7.030(5) |
| *b* (Å) | 8.811(5) |
| *c* (Å) | 7.638(5) |
| β(°) | 114.891(5) |
| Volume (Å$^3$) | 429.2(5) |
| Z | 2 |
| No. Unique Reflections | 1660 |
| No. of parameters | 164 |
| $R[F^2>2\sigma(F^2)]$, $wR_2(F^2)$ | 0.0451, 0.0977 |
| $\Delta\rho_{min}$, $\Delta\rho_{max}$ (eÅ$^{-3}$) | -0.899, 0.0623 |
| Goodness of Fit | 1.183 |

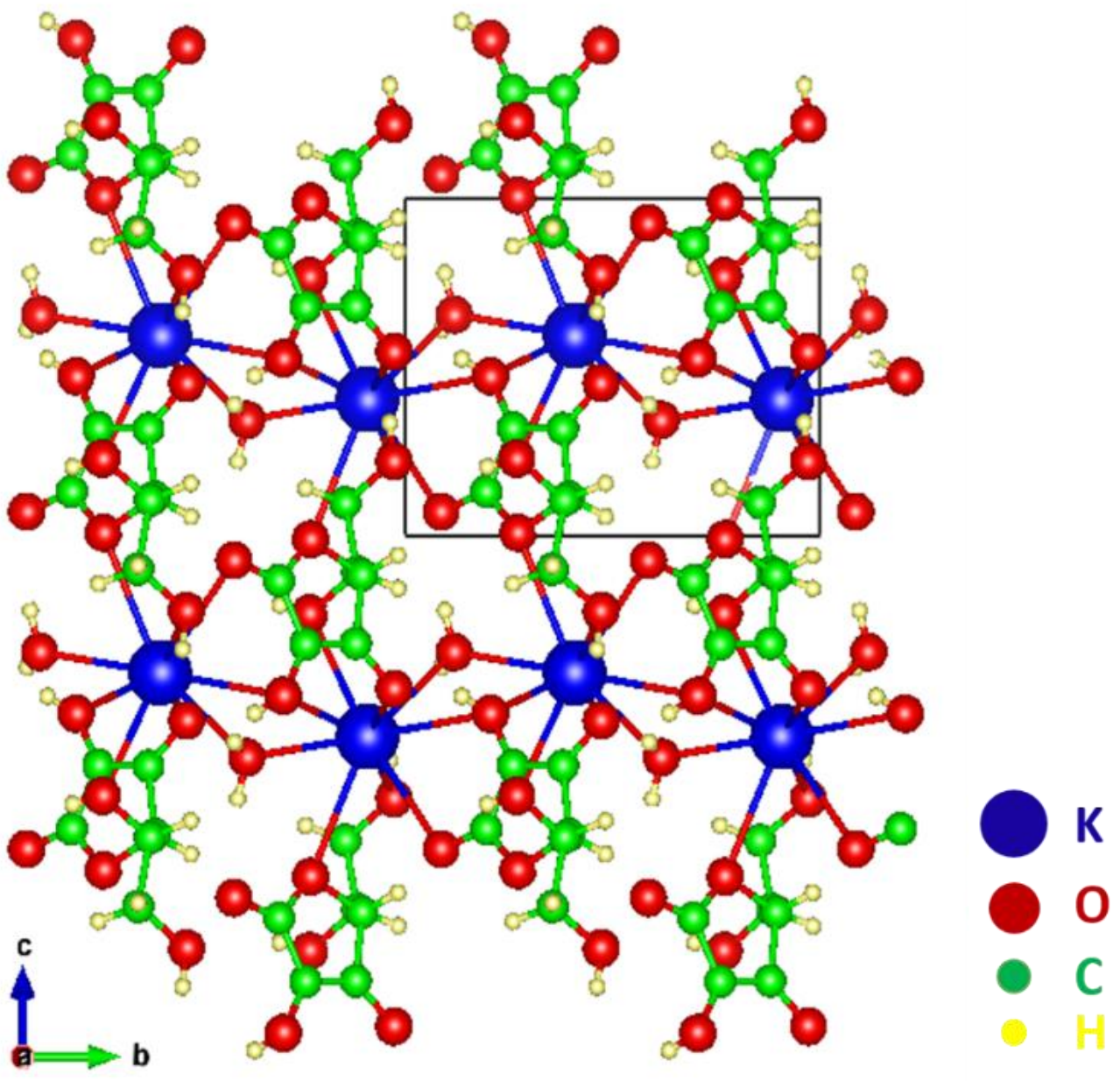

Figure.3. Packing diagram of potassium l-ascorbate monohydrate perpendicular to **a** direction (double bonds are not shown for simplicity).

## Powder XRD

Powder XRD was carried out for checking the single-phase formation in the bulk and for purity. For confirmation of bulk purity, PXRD data were collected for powder from different parts of the crystal. Data were refined by the Rietveld method [27] using Fullprof software [28]. The refinement results are shown in Figure.4. Experimental data and calculated data match well which confirm the presence of single phase.

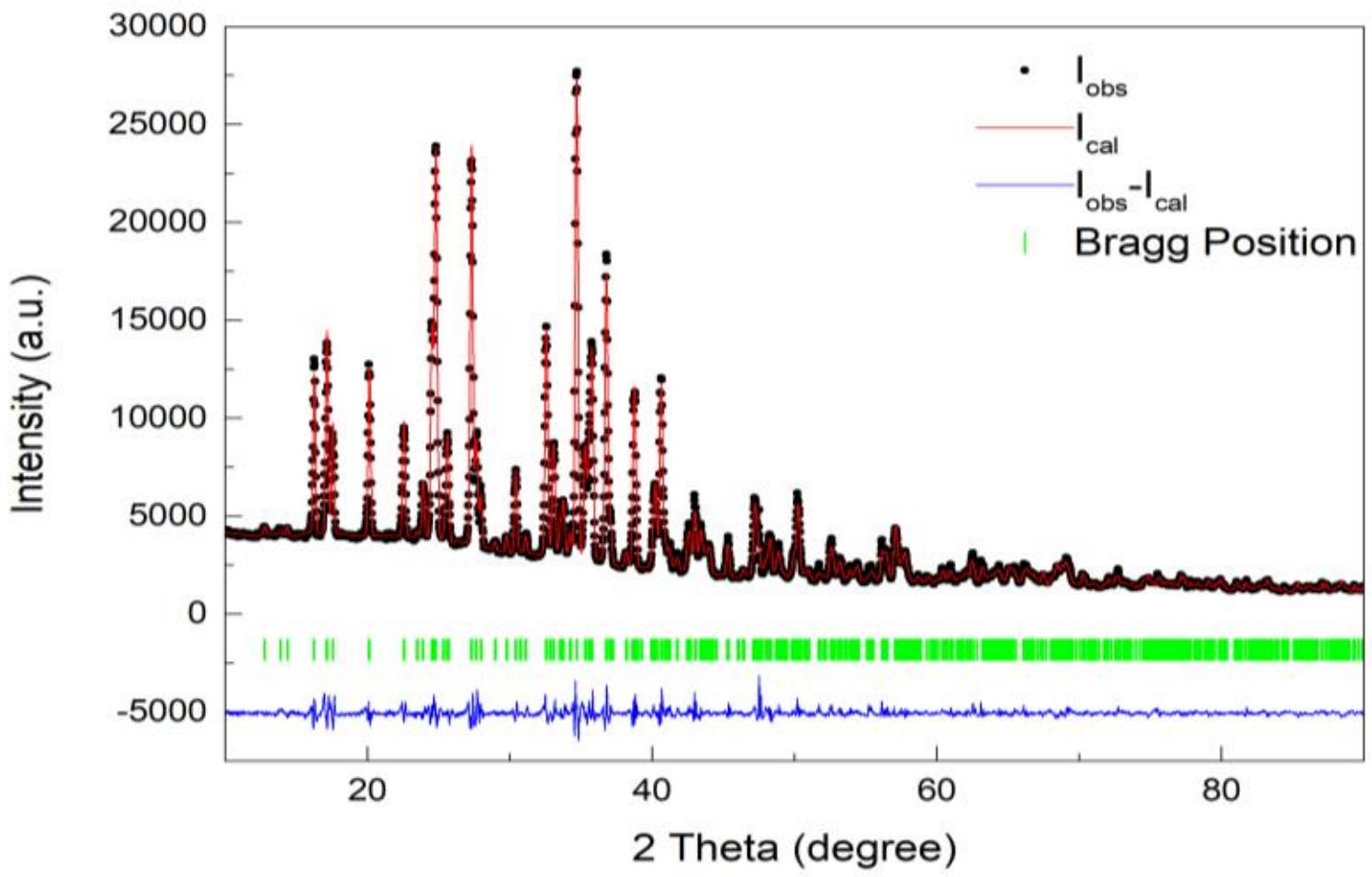

Figure.4. Rietveld refined powder X-ray pattern of KLAM

## Rocking curve measurement: -

Rocking curve study was carried out on a polished a-plate of the crystal. A Gaussian profile fitting was made and FWHM was obtained to be 0.0023° (72 arc-second) which shows that quality of the grown crystal is good. The rocking curve (intensity vs omega) is shown in Figure.5.

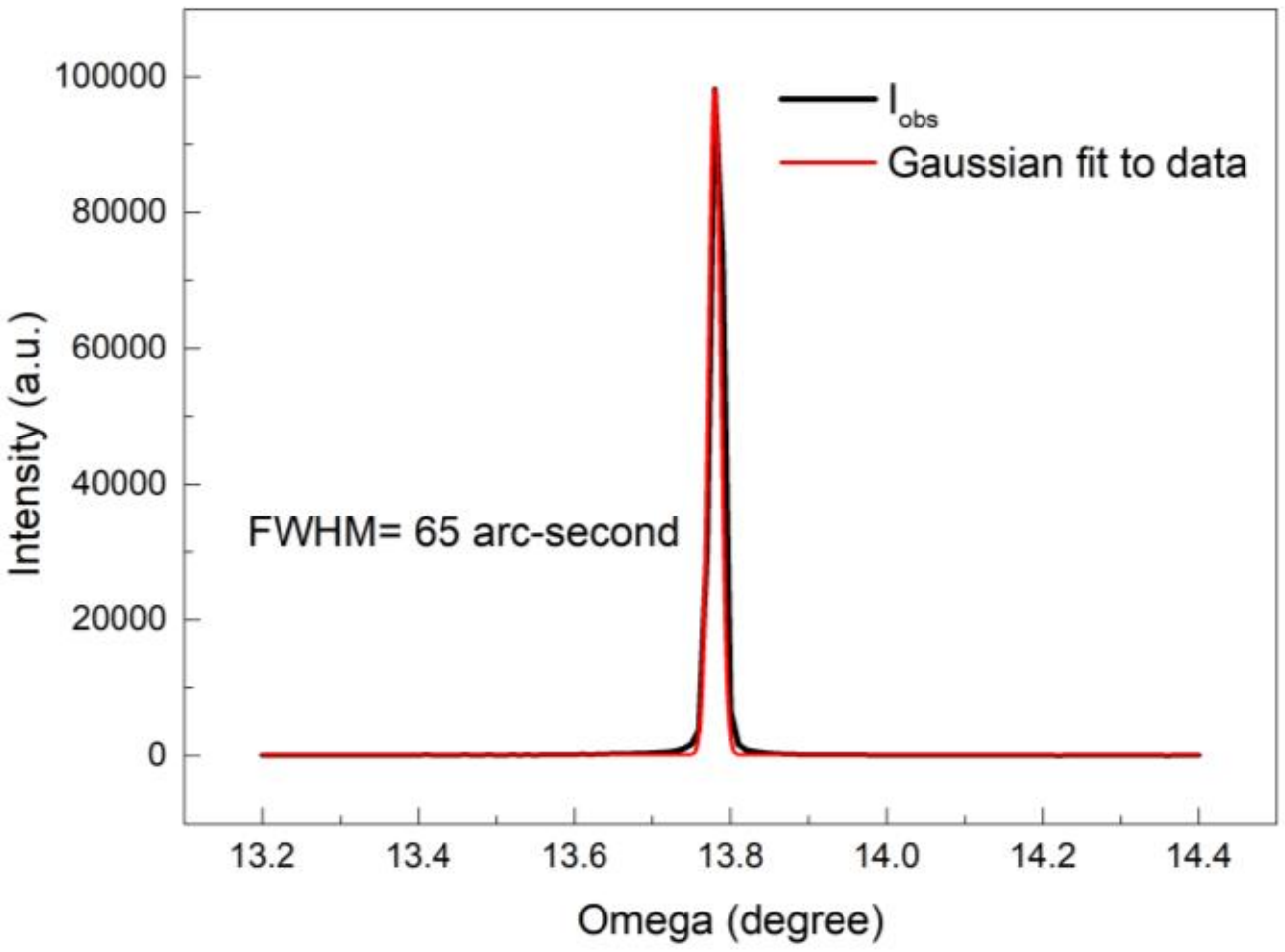

Figure.5. Rocking curve profile measured on **a** plate of KLAM crystal

## Morphology

Morphology of the crystal was generated using WinXmorph software [29,30]. Based on crystallographic data provided the software generates morphology appropriate with point group symmetry. Selection of hkl values and appropriate central distances of the faces yields a pattern identical to the crystal's actual morphology. Angles between the planes calculated from the software and those measured experimentally using a contact goniometer are listed in table II. Theoretical and experimentally measured interplanar angles show a good match. The morphology of the crystal generated is illustrated in Figure.6(a) which looks similar to the grown crystal 6(b). The crystal is bulky with 15 habit faces, though fewer faces develop (~12 number) in smaller crystals.

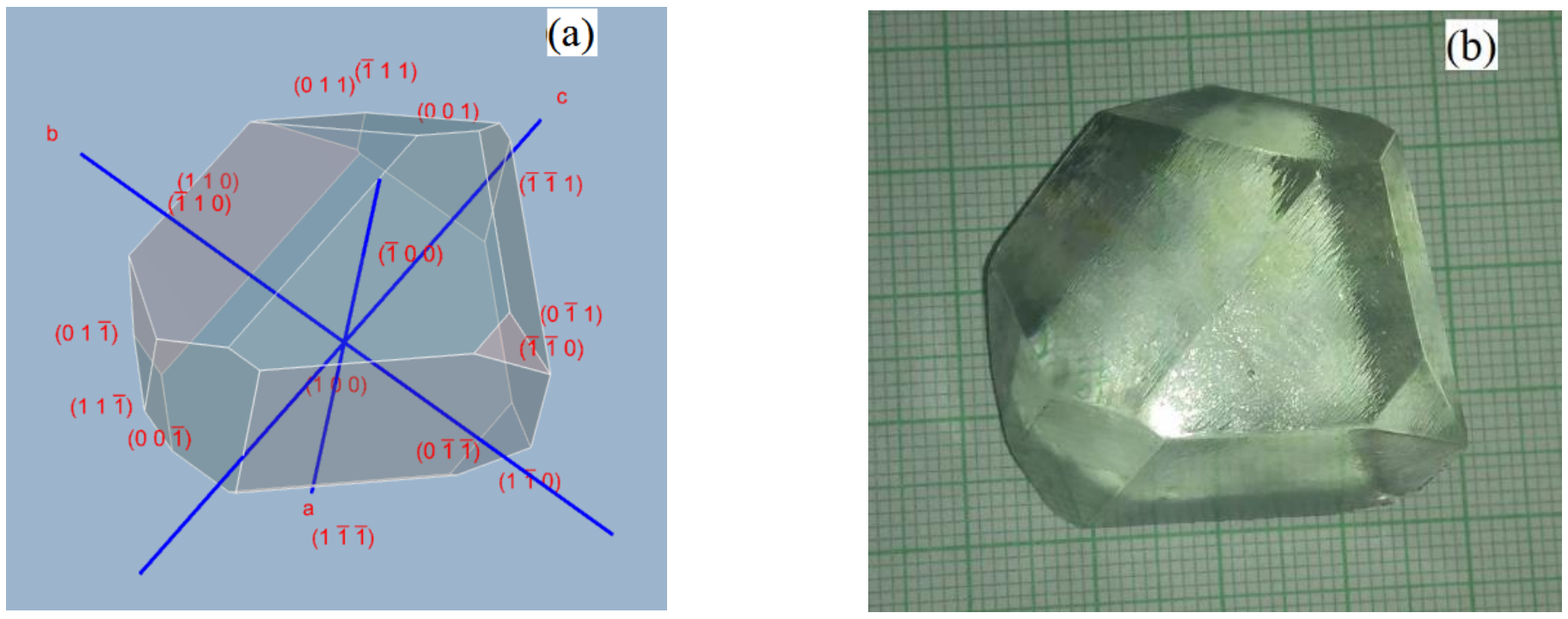

Figure.6. (a) Morphology of the KLAM crystal generated by the Orient Express software which matches with the morphology of the crystal (b)

Table II: Theoretical and experimental values of interplanar angles in KLAM.

| Faces | Interfacial angles in degree | |
|---|---|---|
| | Theoretical | Experimental |
| (100) and (1 1 0) | 35.89 | 35.7 |
| (100) and (0 -1 1) | 70.68 | 70.7 |
| (100) and (0 0 1) | 65.11 | 65.6 |
| (100) and (0 0 -1) | 114.89 | 114.6 |
| (0 -1 -1) and (0 -1 1) | 103.64 | 103.7 |
| (0 -1 -1) and (0 0 -1) | 38.18 | 38.2 |
| (0 -1 -1) and (-1 0 0) | 70.68 | 70.9 |
| (-1 1 1) and ( 0 -1 1) | 92.48 | 92.5 |
| (0 -1 1) and (0 0 1) | 38.18 | 38.25 |
| (-1 0 0 and (-1 1 0) | 35.89 | 35.5 |
| (-1 0 0) and (-1 -1 1) | 60.99 | 61.0 |
| (-1 1 1) and (-1 1 1) | 43.22 | 42.8 |
| (-1 1 0) and (1 1 0) | 108.21 | 107.4 |
| (-1 1 0 ) and (0 0 -1) | 70.06 | 69.7 |
| (-1 1 0) and (0 1 -1) | 50.92 | 50.7 |
| (1 1 0) and (0 0 1) | 70.06 | 70.5 |
| (1 1 0) and (0 1 -1) | 84.58 | 84.5 |

## Thermal analysis

DSC and TGA curves for KLAM are shown in Figure.7. Both measurements were performed at 5 °C/min heating rate. TGA curve shows that the material starts to lose weight from 80 °C. From 80°C to 105 °C degree, it loses ~0.2121 mg weight, which is 5.81% of its initial weight of 3.65 mg. The theoretically calculated percentage of water is 7.751%. This is close to the experimentally measured value and indicates that KLAM loses most of the water between 80 °C to 105°C. The range may vary with different heating rates. Beyond 105°C, all remaining water goes out of the sample. At higher temperatures, weight loss couldn't be measured, as the material becomes foam and bubbles out of the crucible. An endothermic peak at 89 °C in the DSC curve is attributed to dissociation of water molecules from the KLAM and subsequent evaporation of crystallized water, which is also in conformance with TGA data. During the second heating (of the same material), the endothermic peak was not observed; which confirms that the endothermic peak is due to dissociation and evaporation of water molecule from KLAM. Results of both TGA and DSC endorse that KLAM is stable at least up to 80°C which is the onset temperature in DSC. After this temperature, it loses crystallinity.

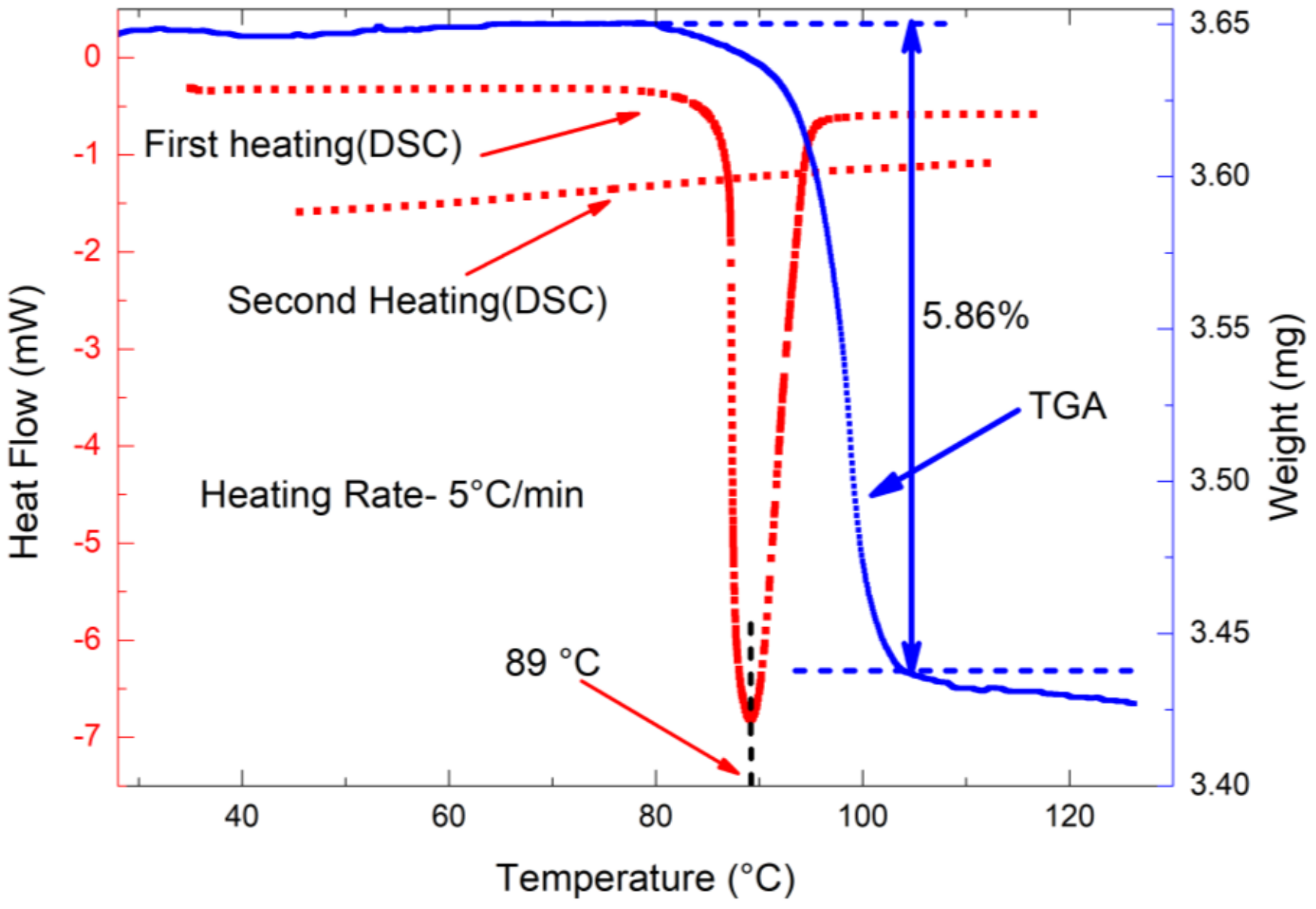

Figure.7. TGA and DSC plots of KLAM. Note that the peak at 89° in the first DSC curve matches the onset of the weight loss in the TGA curve.

## UV-Vis spectroscopy

Figure.8 shows the optical transmission spectrum of KLAM. A polished 1mm thick a-plate was used. The graph shows that crystal has transparency of more than 80% in visible and good transmission in UV regions. After 320 nm transmission drops sharply and crystal becomes opaque at lower wavelengths. The lower cutoff wavelength is 297 nm and the higher cutoff is near 1500 nm, which indicate that crystal can be used in nonlinear optics to obtain harmonics like SHG and THG in visible and UV region. The value of the bandgap calculated from the lower cutoff is 4.18 eV.

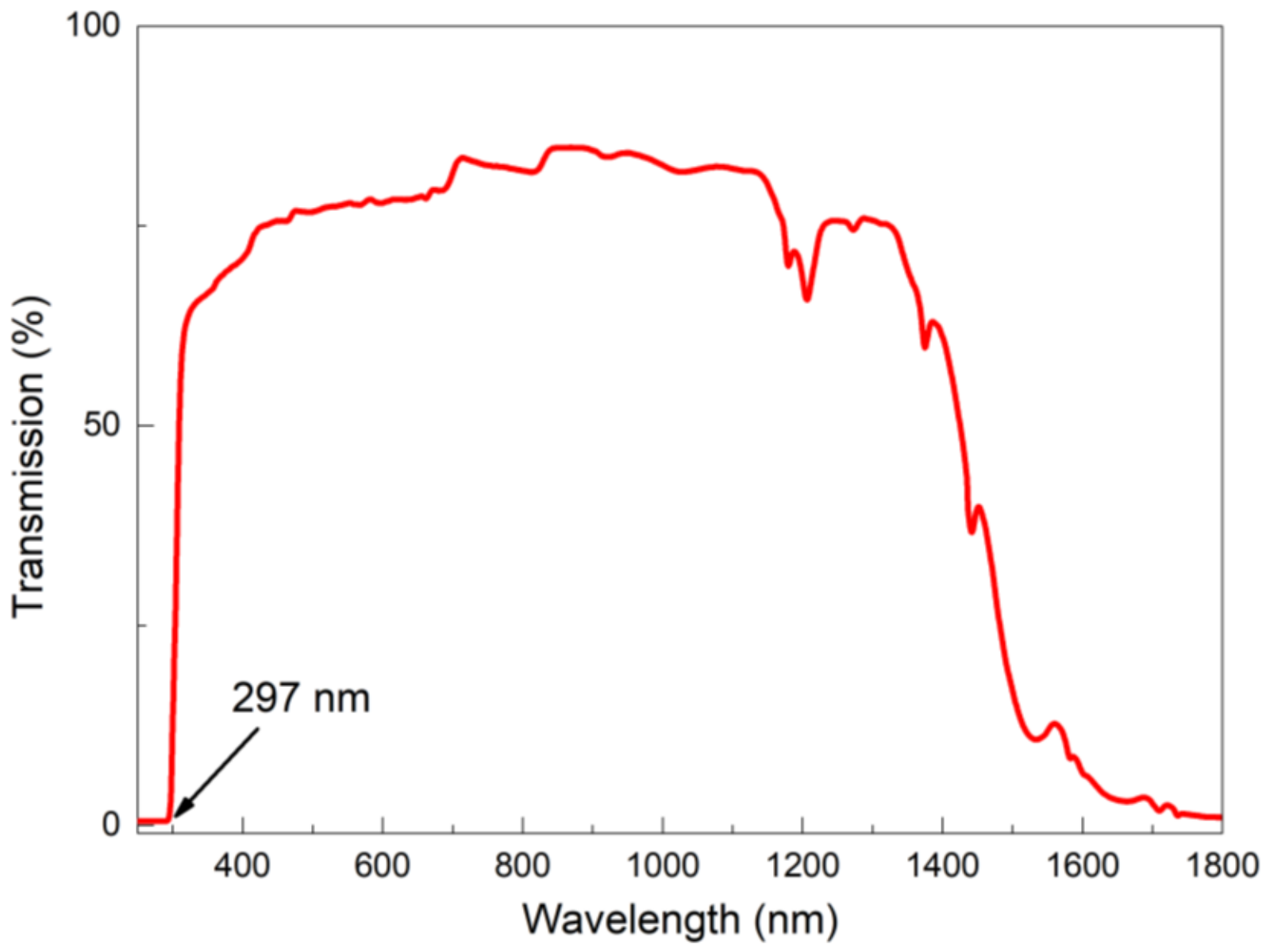

Figure.8. UV-Vis Transmission spectrum of KLAM crystal recorded on a polished **a** plate

## Powder SHG measurements

Since KLAM crystallizes in the non-centrosymmetric space group, it is expected to be NLO-active. We carried out SHG measurements using the Kurtz−Perry method [31]. Here laser beam of wavelength 1064 nm in $TEM_{00}$ mode with pulse duration of 10 ns and repetition rate of 10 Hz was used. The crystal was ground to fine powder and tightly packed in a capillary tube. The capillary tube was mounted in the path of the laser beam. The output light was passed through a monochromator transmitting only the second-harmonic (green) light of wavelength 532 nm. The green light intensity was measured by a photomultiplier tube (PMT) and the ensuring electrical signal was then displayed on the oscilloscope. This procedure was repeated for the standard KDP powder and the ratio of the second-harmonic intensity outputs was calculated. The ratio shows that the second harmonic conversion efficiency of KLAM is 3.5 times that of KDP.

## Collinear and Spontaneous noncollinear phase matching

Spontaneous Noncollinear phase matching (SNCPM), also known as vector phase matching, is commonly observed in crystals with higher birefringence and large nonlinear optical coefficients. NCPM arises due to the interaction of the fundamental beam with scattered light inside the crystal. When fundamental beam with wave vector $k_1(\lambda)$ falls on the crystal, most of it passes through the crystal unaltered but a small amount of the beam is scattered from the crystal surface and crystal imperfections, yielding the wave vector $k_1'(\lambda)$. If the wave vector for the second harmonic wave is $k_2(2\lambda)$, where the phase-matching condition $\Delta k = (k_2 - 2k_1) = 0$ is satisfied, second harmonic light is amplified [32].

To observe SNCPM and collinear phase matching (CPM), **a**-plate of KLAM crystal was oriented such that that crystallographic [100] direction was along the OX axis and [010] direction along the OZ axis of the goniometer. The corresponding Laue pattern (experimental and computed) for this orientation is shown in Figure.9(a). The positive directions of rotation of the goniometer axes are also shown in Figure.9(b).

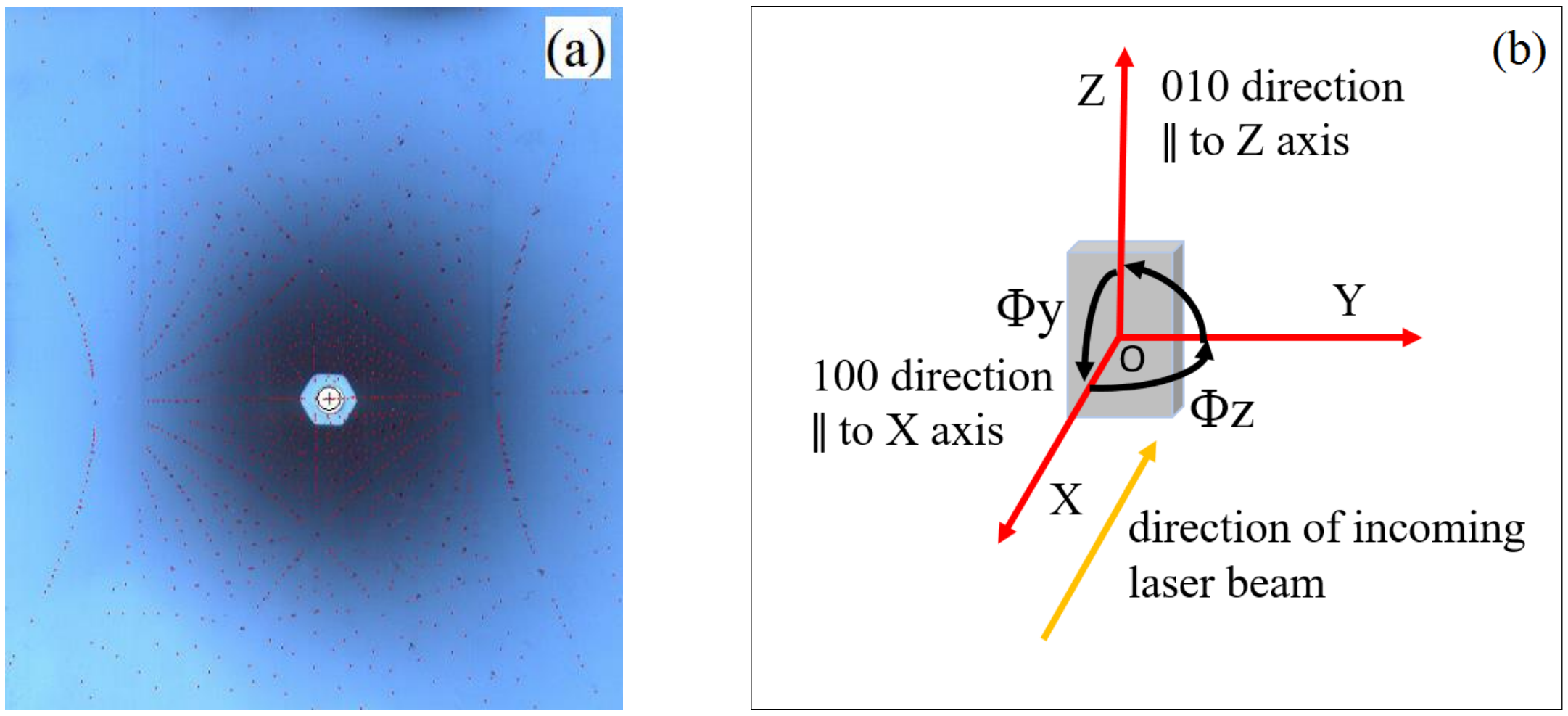

Figure. 9. (a)Fitted Laue pattern in 100 direction. (b)Crystallographic and incoming laser beam direction with respect to goniometer axis. Arrows in black color show the positive direction of rotation.

The crystal was then rotated around [010] direction during which SNCPM rings were initially observed. On further rotating the crystal anticlockwise, collinear phase matching was observed at 13.5°, the evolution of which is shown in figure.10. On further rotation from CPM direction by small-angle (~ 2°), the intensity of SHG signal drastically reduces. Thus, SNCPM has helped us in identifying a phase-matching direction.

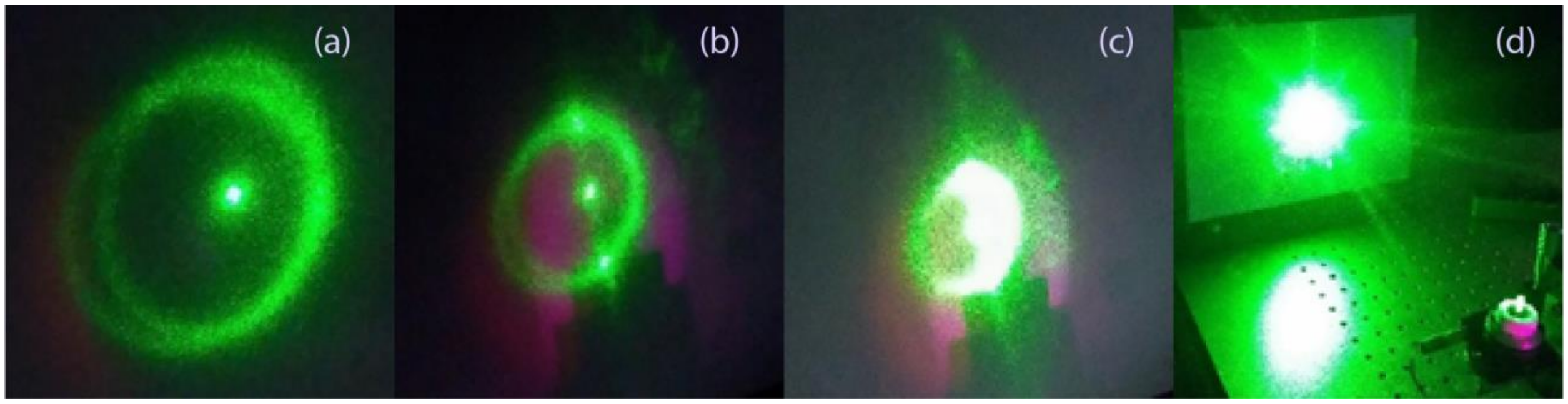

Figure.10. Evolution of SNCPM rings into collinear PM spot in the **a-** plate of KLAM crystal.

To measure the other phase-matching directions experimentally, rotation around OY direction was employed with the rotation around OZ axis. List and 3D plot (Matlab) of angles at which phase matching was obtained by varying Φy and Φz are shown in table III and Figure.11 respectively.

Table III: Values of Φy and Φz for which phase matching directions observed

| Φy (in degree) | Φz (in degree) |
|---|---|
| -21.67 | -3 |
| -19.67 | -5.5 |
| -16.84 | -7 |
| -13 | -9 |
| -10 | -11.5 |
| -7 | -13 |
| -5 | -13 |
| -3 | -13.5 |
| -1 | -14 |
| 0 | -13.5 |
| 3.33 | -13.5 |
| 6.67 | -12 |
| 9.67 | -11 |
| 12.67 | -9 |
| 15.67 | -7 |
| 18.67 | -4 |
| 21.67 | -1 |
| 24.33 | 2 |
| 27.33 | 4.5 |
| 28.33 | 6.5 |

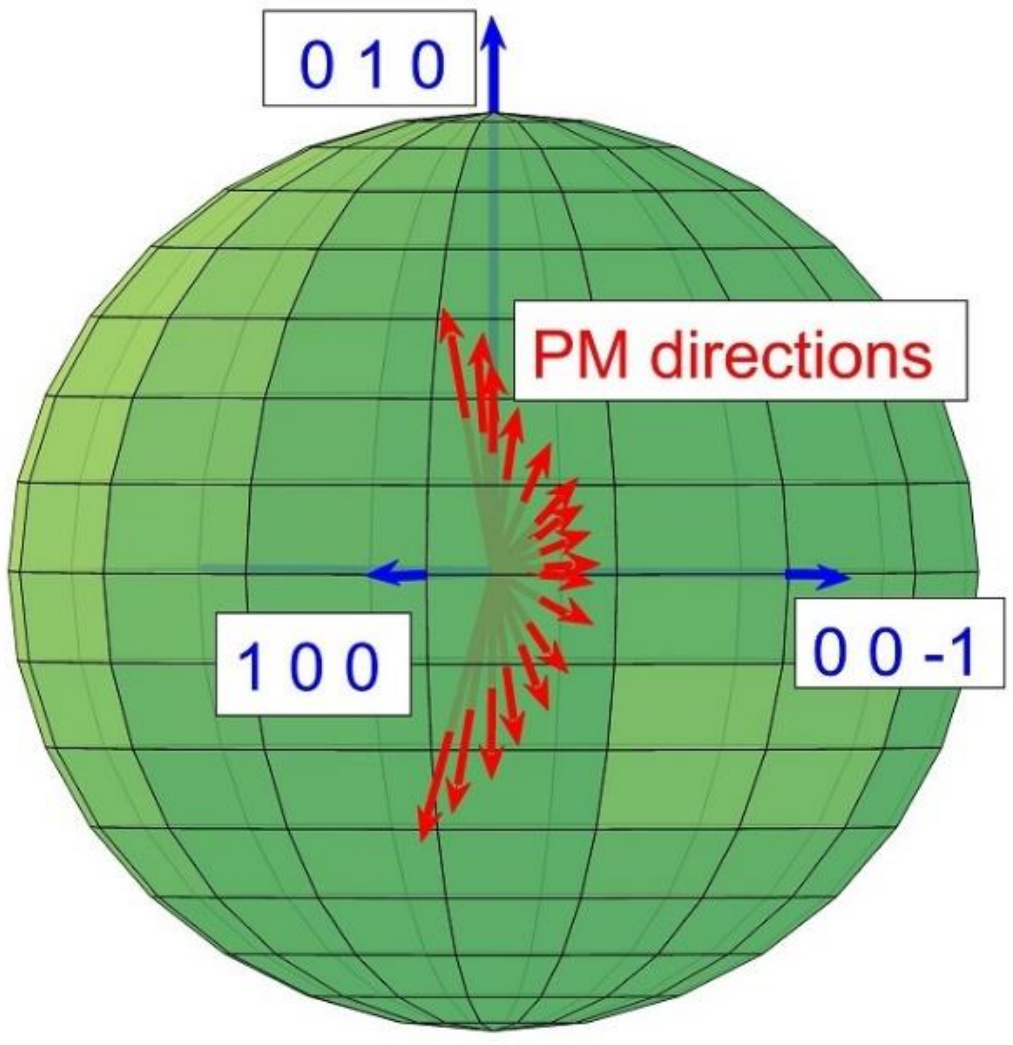

Figure.11. 3D plot of the phase-matching directions.

## Laser damage studies

In nonlinear optics, the crystal's ability to withstand the high power density of laser energy is a vital factor. In laser damage studies single shot (pulse) and multiple shot threshold values are important. In our experiment, the damage threshold is defined as the highest power density value up to which the crystal surface remains undamaged. For every pulse, a fresh area of the surface was exposed to avoid history effect. Single shot (pulse) and multiple shot damage threshold values of KLAM crystal were determined using Nd:YAG laser of 20 ns pulse width and 10 Hz repetition rate in $TEM_{00}$ mode. Multiple shot damage studies were carried out by employing the procedure defined by Nakatani [33]. For multiple shots, the threshold value is always lower than the single-shot value due to cumulative effect [34]. Laser beam is focused on the surface of crystal using a planoconvex lens of focal length 30 cm that was placed at the beam waist. The diameter of the focused spot was calculated using the formula-

$$d'=4M^2\lambda f/\pi d$$

where $M^2$ is the beam quality factor (which has a value equal to 2 for the laser used in this study), $\lambda$ is the wavelength of the laser beam, $f$ is the focal length of the lens and d is the diameter of the laser beam waist. For 6 mm diameter at beam waist, the calculated diameter (d') of spot for 1064 nm was 135 μm. The damage was monitored continuously by observing the He-Ne laser intensity passing collinearly with the Nd:YAG laser through the same spot [25]. The occurrence of damage causes sharp reduction in the intensity of He-Ne laser. Further verification of the damage was done by examining the crystal under a Leitz orthoplan microscope.

A polished **a** plate of the KLAM crystal was used for damage studies. Figure.12 shows the graph of the number of pulses and average power density at which the damage occurred. The average damage threshold value for single pulse is 3.07 GW/cm$^2$ and for multiple pulses (2000 pulses) is 2.85 GW/cm$^2$. The cause of damage is inferred as dielectric breakdown as for short pulses ($10^{-8}$ to$10^{-10}$ ns pulse width) this is the most prominent reason [35]. There was no sign of melting after the damage which supports this view. Nearly 70% of the damage was observed on the exit surface of the crystal. Boling and Crisp [36] brought forward an elegant and simple explanation for the electric field associated with the beam at the entrance and exit surfaces. According to this, the electric field is higher at the exit surface because the Fresnel reflected field at the entrance surface is 180° out of phase with the incident field, whereas at the exit surface, the reflected field is in phase with the incident field. Thus, the net electric field is higher at the exit surface than at the entrance surface for a given input irradiance.

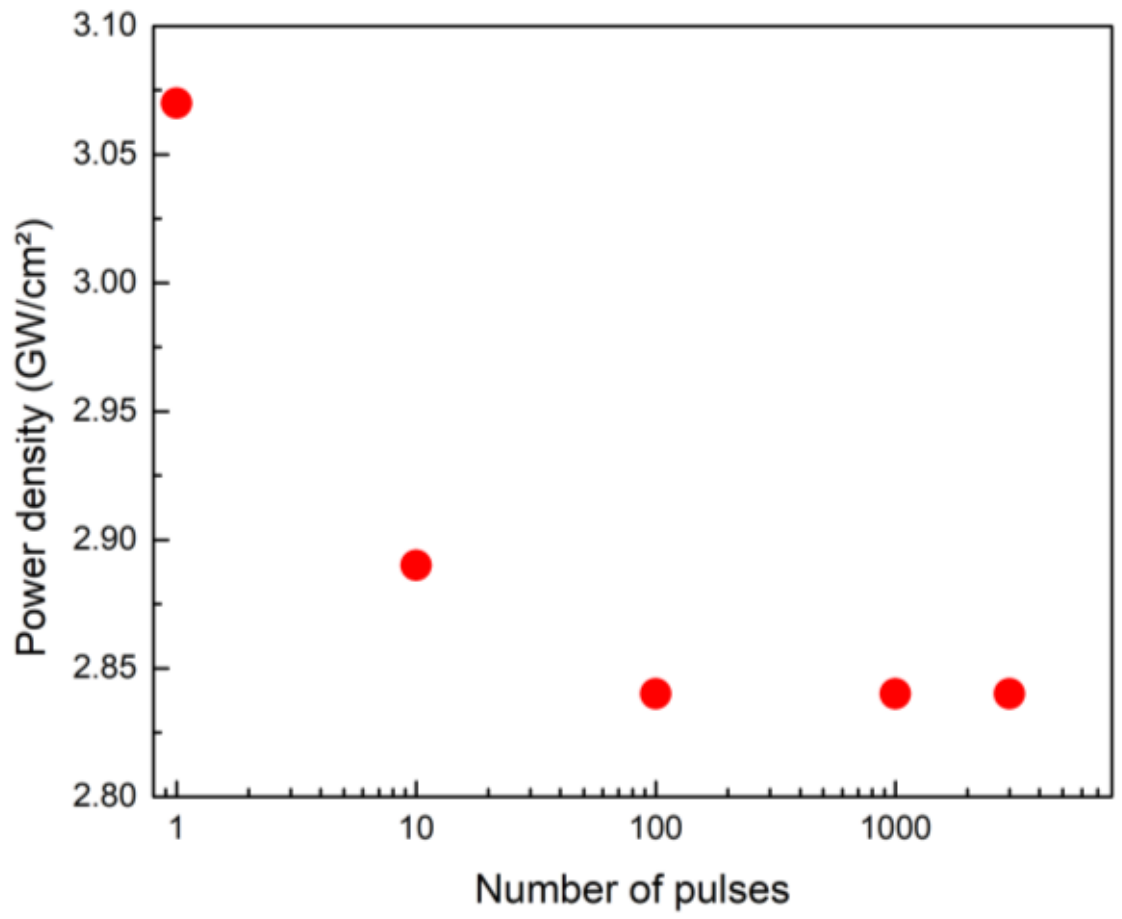

Figure.12: Damage threshold (average) value as a function of the number of pulses.

## Conclusion

Large size crystals of KLAM were successfully grown by solution growth technique. This material is non-toxic and does not pose any danger to the environment. Single crystal XRD reveals that KLAM crystallizes in the monoclinic $P2_1$ space group. The crystal possesses bulky morphology and is stable up to 80 °C after which it loses water and hence crystallinity. SHG conversion efficiency of this material is 3.5 times that of KDP, as revealed by powder Kurtz Perry technique. In noncollinear phase matching, rings up to third order were observed which indicate high conversion efficiency and large birefringence. Noncollinear phase matching was observed on **b** and **c** plates. Collinear phase matching was experimentally observed on **a** plate of the crystal. Laser damage threshold value observed on **a** plate is 3.07 GW/cm$^2$. Good SHG conversion efficiency, collinear and spontaneous noncollinear phase matching ability make this material potentially useful for nonlinear optical applications.

## Acknowledgement

The authors are grateful to Prof. P.K. Das, IISc., Bangalore for the measurement of SHG efficiency by Kurtz-Perry method. The authors acknowledge the support of the departmental central facility funded by the University Grants Commission.

## Bibliography

[1] T. Taira and T. Kobayashi: IEEE J. Quantum Electron. 30, 800 (1994).
[2] G.A. Rines, H.H. Zenzie, R.A. Schwarz, Y. Isyanova, and P.F. Moulton, IEEE J. of Selected Topics in Quantum Electron. 1, 50 (1995).
[3] P.V. Kolinsky, Opt.Eng. 31, 1676 (1992).
[4] H. S. Nalwa, App. Organomet. Chem. 5, 349 (1991).
[5] P. A. Franken and J. F. Ward, Rev. Mod. Phys. 35, 23 (1963).
[6] P. A. Franken, A. E. Hill, C. W. Peters and G. Weinreich, Phys. Rev. Lett. 7, 118 (1961).
[7] A. Giordmaine, Phys. Rev. Letters 8, 19 (1962).
[8] C. C. Wang, and G. W. Racette, Appl. Phys. Lett. 6, 169 (1965).
[9] J. E. Bjorkholm, Phys. Rev. 142, 126 (1966).
[10] N Bloembergen and P. S. Pershan, Phys. Rev. 128, 606(1962).
[11] J. A. Giordmaine, and, R. C. Miller, Phys. Rev. Lett. 14, 973 (1965).
[12] J. F. Ward and P. A. Franken, Phys. Rev. 133, A183 (1964).
[13] Y. R. SHEN, The Principles of Nonlinear Optics, Wiley, New York (1984).
[14] R. W. Boyd, Nonlinear Optics (Academic, Boston, 1992).
[15] Y.N. Xia, C.T. Chen, D.Y. Tang, B.C. Wu, Adv. Mater. 7, 79 (1995).
[16] L. Kang, S. Y. Luo, H.W. Huang, N. Ye, Z.S. Lin, J.G. Qin, C.T. Chen, J. Phy. Chem. C 117, 25684 (2013).
[17] T. T. Tran, H. W. Yu, J. M. Rondinelli, K. R. Poeppelmeier, P. S. Halasyamani, Chem. Mater. 28, 5238 (2016).
[18] J. Hvoslef, Acta Cryst., B24, 1431 (1968).
[19] V Kuellmer, Kirk-Othmer Encyclopedia of Chemical Technology 25, 17 (2000).
[20] G. Bourhill, K. Mansour, K.J. Perry, L. Khundkar, E. T. Sleva, R. Kern, and J. W. Perry, I. D. Williams, S. K. Kurtz, Chem. Mater. 5, 802 (1993).
[21] K. Raghavendra Rao, C. Aneesh, H.L. Bhat, S. Elizabeth, M.S. Pavan, T.N. Guru Row, Cryst. Growth Des. 13, 97 (2013)
[22] K. Raghavendra Rao, H.L. Bhat, S. Elizabeth, CrystEngComm 15, 6594 (2013)
[23] K. Raghavendra Rao, H.L. Bhat, S. Elizabeth, Mater. Chem. Phys. 137, 756 (2013)

[24] K. Raghavendra Rao, R. Sanathkumar, H. L. Bhat, S. Elizabeth, Applied Physics B, 122 (2016) 270.
[25] R. Austria, A. Semenzato, A. Bettero, J Pharm Biomed Anal 15, 795 (1997).
[26] A.E. Kellie, S.S. Zilva, Biochem. J. 32, 1561 (1938).
[27] H. M. Rietveld, J. Appl. Cryst. 2, 65 (1969).
[28] J. Rodríguez-Carvajal, FullProf Suite. Available at http://www.ill.eu/sites/fullprof/index.html.
[29] W. Kaminsky, J. Appl. Crystallogr. 38, 566 (2005). http://cad4.cpac.washington.edu/WinXMorphHome/WinXMorph.htm.
[30] W.J. Kaminsky, J. Appl. Cryst. 40, 382 (2007).
[31] S. K. Kurtz and T. T. Perry, Journal of Applied Physics 39, 3798 (1968).
[32] J.A. Giordmaine, Phys. Rev. Lett. 8, 19 (1962).
[33] H. Nakatani, W.R. Bosenberg, L.K. Cheng, C.L. Tang, Appl. Phys. Lett. 53, 2587 (1988)
[34] V. Venkataramanan, C. K. Subramanian and H. L. Bhat, J.Appl. Phys. 77, 6049 (1995).
[35] R. M. Wood, Laser-Induced Damage of Optical Materials (CRC Press, 2003).
[36] N.L. Boling, M.D. Crisp, G. Dube, Appl. Opt. 12, 650 (1973).